# Quaternion optical computing chip for parallel high-dimensional data processing

Songyue Liu[1], Qi Lu[1], Yuan Zhong[1], Yuru Li[2], Meng Xiang[3], Zhaohui Li[4], Chao Lu[5], Yikai Su[1*], and Lu Sun[1*]

[1]State Key Lab of Photonics and Communications, Department of Electronic Engineering, Shanghai Jiao Tong University, Shanghai 200240, China

[2]School of Microelectronics Science and Technology, Sun Yat-Sen University, Zhuhai 519000, China

[3]Key Laboratory of Photonic Technology for Integrated Sensing and Communication, Ministry of Education, Guangdong University of Technology, Guangzhou 510006, China

[4]Southern Marine Science and Engineering Guangdong Laboratory (Zhuhai), Zhuhai 519000, China

[5]Photonics Research Institute, Department of Electronic and Information Engineering, The Hong Kong Polytechnic University, Hong Kong SAR, China

*Corresponding author: Yikai Su and Lu Sun (email: yikaisu@sjtu.edu.cn; sunlu@sjtu.edu.cn)

## Abstract

Optical computing chips have emerged as a transformative computing technology due to their high computational density, low energy consumption, and compact footprint. While real- and complex-valued computing chips have been well developed, their fundamental limitations in representing high-dimensional data significantly constrain their applicability in modern signal processing. Quaternions enable direct operations on three- and four-dimensional data, powering high-dimensional processing in data analytics and artificial intelligence. Here we demonstrate a quaternion optical computing chip (QOCC) for the first time and benchmark its performance in several typical application scenarios: three-dimensional point cloud processing, RGB chromatic transformation, and quaternion convolutional neural network for color image recognition. The QOCC harnesses high parallelism of light by wavelength-division multiplexing, processing high-dimensional data simultaneously through multiple optical wavelength channels. Compared to the electronic computing counterpart, our QOCC achieves higher computational fidelity (root mean square error < 0.035) and substantially reduced computational load (2/3 lower). It paves the way towards next-generation optical computing, overcoming the limitations of traditional computing systems in high-dimensional data processing.

## Introduction

Artificial intelligence (AI) has witnessed remarkable success in reshaping various aspects of modern life. The training and inference processes of AI impose enormous computational demands on computing chips. With real- and complex-valued computing chips, AI has facilitated significant advances in tasks ranging from pattern recognition[1-5], speech processing[6-10], and decision-making[11-13] to radar[14-16], wireless communications[17-21], and robotics[22,23]. However, these computing architectures remain constrained to low-dimensional representations of data, limiting their application in tensor processing such as three-dimensional (3D) point cloud reconstruction and color image recognition. Hypercomplex algebra, especially quaternions, offers exceptional advantages in handling high-dimensional-data-based tasks. Quaternion computation has been widely used in modern signal processing for its capability of naturally representing four-dimensional data, significantly reducing the number of consecutive multiplications during processing, and robustly preserving the correlations across different dimensions of data. Recent advances in quaternion-valued neural networks are revolutionizing various aspects of our information technology including rotational dynamics modeling, multi-channel data fusion, and high-dimensional feature extraction in fields like computer vision[24,25], autonomous navigation[26,27], advanced signal processing[28,29], etc.

Although quaternion computation can process high-dimensional data more efficiently, it still requires extensive multiply-accumulate (MAC) operations and frequent data access, imposing a substantial computational burden on existing electronic computing hardware. Recently, silicon photonic integrated circuits have emerged as a promising platform for on-chip optical computing and neural network acceleration[30-44], providing unparalleled advantages in bandwidth, parallelism[40] and energy efficiency[43,44]. While real- and complex-valued computing chips have been implemented using micro-ring resonators[33,45] (MRRs) or Mach-Zehnder interferometer arrays[46,47], the development of on-chip quaternion computing remains largely unexplored. New optical computing paradigms and architectures must be developed to achieve high computational precision and efficient acceleration for diverse quaternion-valued computations and neural networks. Bridging the gap between silicon photonics and quaternion computation is essential for high-performance photonic processors capable of parallel high-dimensional data processing.

In this work, we present the first demonstration of a quaternion optical computing chip (QOCC). The chip integrates all key computational functions including input preparation, weight multiplication and output generation on a silicon photonic platform. Different from traditional electronic implementations where cumbersome arithmetic operations are executed sequentially, the QOCC improves parallelism by leveraging wavelength-division multiplexing (WDM) and exploits the inherently low latency of light to achieve high computing speed. Moreover, we showcase its feasibility in several high-dimensional data processing applications such as 3D point cloud reconstruction, RGB chromatic transformation and color image recognition. In all these tasks, our QOCC exhibits high computational fidelity (root mean square error (RMSE) < 0.035) and low computational load (1/3 compared to electronic processors). It also features a 9.2-Gbaud data transmission throughput and a 331.2-GOPS computing speed, significantly surpassing conventional electronic quaternion computing chips. Our work opens new avenues for a wide variety of high-impact and real-time applications ranging from autonomous vehicle perception[48-50] to high-precision object detection[51-55] and large-scale drone swarm coordination[56-60].

# Results

## Principle of operation

Quaternions, an extension of complex numbers first introduced by Hamilton, are widely used to represent and process high-dimensional data. The unique algebraic properties of quaternions include but are not limited to compact representation of rotations as a single operation, avoidance of gimbal lock[61], efficient composition of quaternion components via multiplication, support for smooth interpolation[62], and inherent preservation of inter-dimensional correlations. Among them, the ability to preserve correlations between different dimensions of data is crucial for performing tasks like 3D rotation and color image processing. For example, conventional real- and complex-valued approaches process the red, green and blue channels of a color image independently, thereby overlooking the information embedded in their inter-channel correlations. Quaternion-based computations, however, exploit this kind of information to achieve higher accuracy in color image recognition. Therefore, they find wide applications in many modern real-time control and advanced signal processing systems, including attitude control of drones, autonomous driving through multi-sensor fusion and color image recognition in dynamic environments, as schematically illustrated in Fig. 1a. Obviously, metrics such as computational efficiency, processing latency, numerical stability and arithmetic precision are of central importance for quaternion computing chips tackling these tasks.

Mathematically, a quaternion $q$ can be expressed as $q = q_0 + q_1\mathbf{i} + q_2\mathbf{j} + q_3\mathbf{k}$, with imaginary unit $\mathbf{i}, \mathbf{j}, \mathbf{k}$ satisfying: $\mathbf{i}^2 = \mathbf{j}^2 = \mathbf{k}^2 = \mathbf{ijk} = -1$, $\mathbf{ij} = -\mathbf{ji} = \mathbf{k}$. For any point in 3D space, it can be written as a quaternion $p = (0, \mathbf{p})$, where $\mathbf{p}$ is the position vector. Hereafter, unless otherwise clarified, the italic letters stand for quaternion quantities and the bold letters stand for vectors or matrices. The transformation on the point such as rotation and scaling (Fig. 1b) can be represented by a quaternion operator $q$. The point $p'$ after transformation $q$ is related to the original point $p$ by $p' = qpq^{-1}$. Such an operation can also be expressed using a transformation matrix $\mathbf{R}(q)$ (see Supplementary Note 1 for more details)

$$\mathbf{R}(q) = \begin{bmatrix} 1-2q_2^2-2q_3^2 & 2q_1q_2-2q_0q_3 & 2q_0q_3+2q_0q_2 \\ 2q_1q_2+2q_0q_3 & 1-2q_1^2-2q_3^2 & 2q_2q_3-2q_0q_1 \\ 2q_1q_3-2q_0q_2 & 2q_2q_3+2q_0q_1 & 1-2q_1^2-2q_2^2 \end{bmatrix} = \begin{bmatrix} w_{11} & w_{12} & w_{13} \\ w_{21} & w_{22} & w_{23} \\ w_{31} & w_{32} & w_{33} \end{bmatrix}, \quad (1)$$

where the rightmost matrix in Eq. (1) is the weight matrix $\mathbf{w}$ we use to configure the QOCC. The experimental setup for performing the quaternion computations and accelerating the quaternion convolutional neural networks (QCNN) based on the QOCC is shown in Fig. 1c. The 3D signal represented by the quaternion $p = 0 + x_i\mathbf{i} + y_i\mathbf{j} + z_i\mathbf{k}$ has its components $x_i$, $y_i$, and $z_i$ assigned to optical carriers with equal power but different wavelengths ($\lambda_1, \lambda_2, \lambda_3$). The multi-wavelength light is then multiplexed using a dense wavelength division multiplexer (DWDM). Signal encoding is implemented via an arbitrary waveform generator (AWG), which maps the coordinates of $n$ points to a sequence $\mathbf{S_i} = [x_{i1}, y_{i1}, z_{i1}, x_{i2}, y_{i2}, z_{i2} \ldots x_{in}, y_{in}, z_{in}]$. The encoded electrical signal is amplified by an electrical amplifier (EA) and then fed into a Mach-Zehnder modulator (MZM) to perform high-speed electro-optic modulation on the triple-wavelength-multiplexed optical signal, thereby producing three encoded replicas of dataset $\mathbf{S_i}$. The light wave is then transmitted through optical time delay line (OTDL) with a delay step (between adjacent wavelengths) equal to the symbol duration of $S_i$, effectively achieving time and wavelength interleaving. The time delay $\Delta t$ is set to be equal to the reciprocal of the baud rate of the modulation signal $f_b$ (i.e., $\Delta t = 1/f_b$), which ensures precise temporal alignment of the signals encoded on three wavelengths after synchronizing their periods. The time-and-wavelength-interleaved optical signals are then coupled into the QOCC for quaternion processing. The proposed QOCC comprises three MRR branches, each integrating three thermally tunable MRRs in the add-drop configuration (The design parameters of the QOCC are detailed in Supplementary Note 2). Through precise control of applied power to individual microheaters, the operational weight $w_{ij}$ of each micro-ring within the QOCC can be dynamically configured. The output optical signals from the QOCC are converted back to electrical signals via a balanced photodetector (BPD), accomplishing the MAC operations with high parallelism. Some redundant data is contained in the output sequence of QOCC and should be eliminated before performing following processing such as feature extraction. Dividing the sequence into subsets containing three time slots, only the data in the third time slots contributes to the final results of quaternion computations, as indicated by the black dashed box in Fig. 1c (see Supplementary Note 3 for details). Therefore, we retain the data in the third time slots and output it as $S_o = [(x_{o1}, y_{o1}, z_{o1}), (x_{o1}, y_{o1}, z_{o1}) \ldots (x_{on}, y_{on}, z_{on})]$.

Figure 1d and 1e shows the microscope images of the fabricated devices and the QOCC after packaging. The optical signals carried on three wavelengths propagate through the MRR array simultaneously and experience distinct weighting factors. The transmitted signals are then superposed incoherently through the BPDs, realizing parallel photonic computations. The weighting factors across [-1,1] can be configured by tuning the MRR banks. For example, weights of 1, 1/2, 0, -1/2 and 1 are obtained by controlling the detuning between the working wavelength and the resonant wavelength of the MRR, as indicated by the dashed lines in Fig. 1f (see Supplementary Note 2 for details).

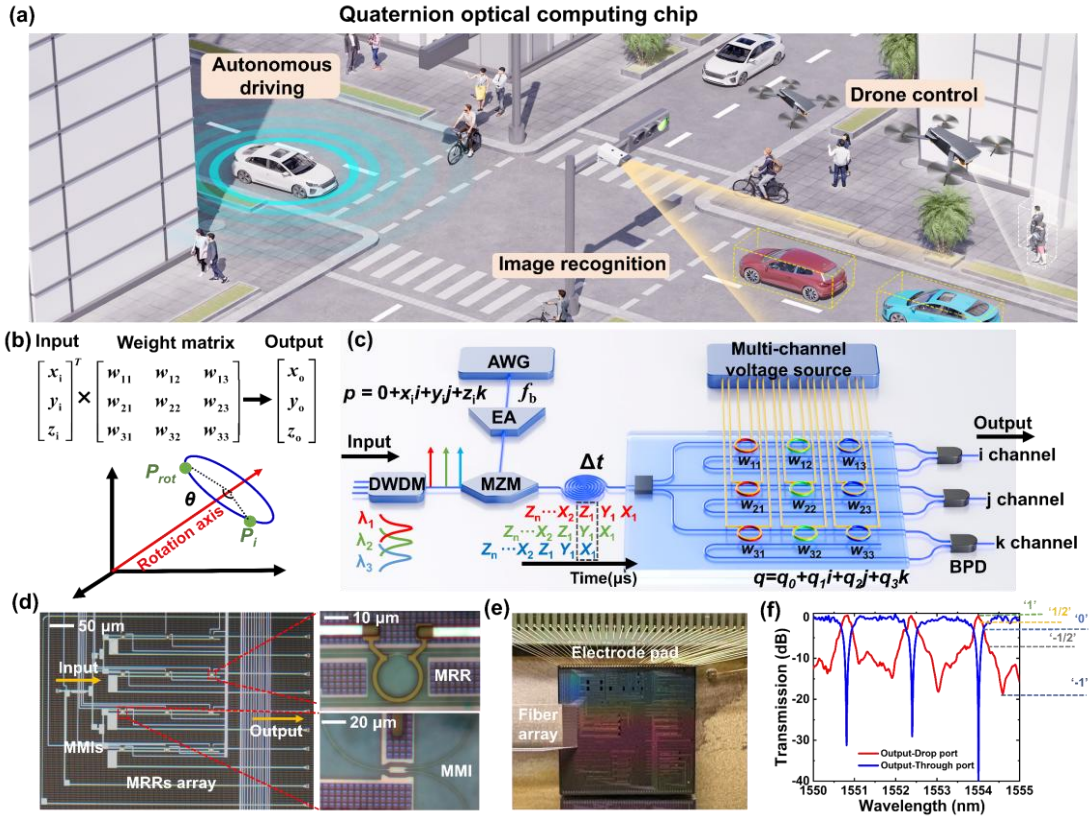

**Fig. 1| QOCC empowering high-dimensional signal processing applications. a,** Typical applications of QOCC, including map reconstruction in autonomous driving, color image recognition in complex environments, and real-time drone control. **b,** Operation on a point in 3D space represented by a quaternion and its equivalent matrix form. **c,** Schematic of the experimental setup for quaternion computations. DWDM: dense wavelength division multiplexer. AWG: arbitrary waveform generator. MZM: Mach-Zehnder modulator. EA: electrical amplifier. BPD: balanced photodetector. **d,e,** Microscope images of the on-chip devices (**d**) and the chip after optical and electrical packaging (**e**). **f,** Weight configuration of the MRRs across [-1,1] through resonance detuning.

### 3D point cloud processing

3D point clouds provide a discrete representation of the geometric structure of objects by sampling spatial coordinates in 3D space. They are widely used across various fields including robotics, autonomous driving, remote sensing, virtual and augmented reality, etc. Transforming the 3D point cloud of an object while preserving its structural integrity demands advanced computational models and hardware accelerators to ensure high fidelity, efficiency, and scalability. Here we perform the 3D point cloud processing using the QOCC. The experimental setup is shown in Fig. 2a. A laser array generated three incoherent optical carriers at wavelengths of 1550.8 nm, 1552.4 nm, and 1554 nm, which were multiplexed by a dense DWDM before being fed into a MZM for electro-optic conversion. During signal encoding, the spatial coordinates of Point 1 - Point $n$ in a 3D point cloud were serialized into a sequence $\mathbf{S} = [x_1, y_1, z_1, x_2, y_2, z_2 \ldots x_n, y_n, z_n]$ and subsequently input to the AWG. The AWG operated at a frequency of 9.2 GHz with a pulse width of 108 ps per data point. Here, we selected a 8-km OTDL with 17 ps/(km·nm) dispersion to create a 1-bit time delay ($\Delta t$ = 108 ps) between signals with adjacent working wavelengths. The pulsed optical signals were then coupled into the QOCC for quaternion computations. Finally, the computation results were obtained by extracting every third bit in the output sequence, as described in detail in Supplementary Note 3.

In this work, we demonstrate the 3D point cloud processing with two models: a 2500-point "SJTU" letter model and a 10000-point "Dog" model. The transformations on the "SJTU" letter model and the "Dog" model are described by the quaternions $q = 1 - 0.4\mathbf{i} - 0.1\mathbf{j} + 0.3\mathbf{k}$ and $q = 1 + 0\mathbf{i} + 0\mathbf{j} - 1\mathbf{k}$, respectively. Since the quaternions are non-unitary (i. e., $|q| \neq 1$), it means the transformations include both rotation and scaling operations. Figure 2b shows the experimental results for the "SJTU" letter model. The left panel displays the original 3D point cloud. The ideal (red curve) and experimental (blue curve) waveforms for the $x, y, z$ coordinates of the point clouds are compared in the middle panel, showing high consistency. In the right panel, we display the ideal and experimental point clouds after transformation in 3D space. As one can see, the morphology of the point cloud processed using the QOCC shows a strong resemblance to the computational prediction. Figure 2c presents the experimental results for the "Dog" model. Compared to the "SJTU" letter model, the "Dog" model has a higher point density, showing finer morphological details (left panel). The waveforms for the ideal (red curve) and experimental (blue curve) point clouds after transformation are shown in the middle panel while the transformed point clouds themselves are displayed in the right panel, demonstrating high fidelity of the QOCC processing. We also carried out a continuous transformation from $q = 1 + 0\mathbf{i} + 0\mathbf{j} + 0\mathbf{k}$ (identity operation) to $q = 1 - 0.4\mathbf{i} - 0.1\mathbf{j} + 0.3\mathbf{k}$ for the "SJTU" letter model and a continuous transformation from $q = 1 + 0\mathbf{i} + 0\mathbf{j} + 0\mathbf{k}$ (identity operation) to $q = 1 + 0\mathbf{i} + 0\mathbf{j} - 1\mathbf{k}$ for the "Dog" model. The snapshots of the 3D point clouds during the transformations can be found in Supplementary Note 4. For both models, the RMSEs between the theoretical and the measured results are less than 0.035, and the positioning accuracy of the QOCC reaches a precision of 5 bit, proving the feasibility of the QOCC in 3D point cloud processing (see details in Supplementary Note 5).

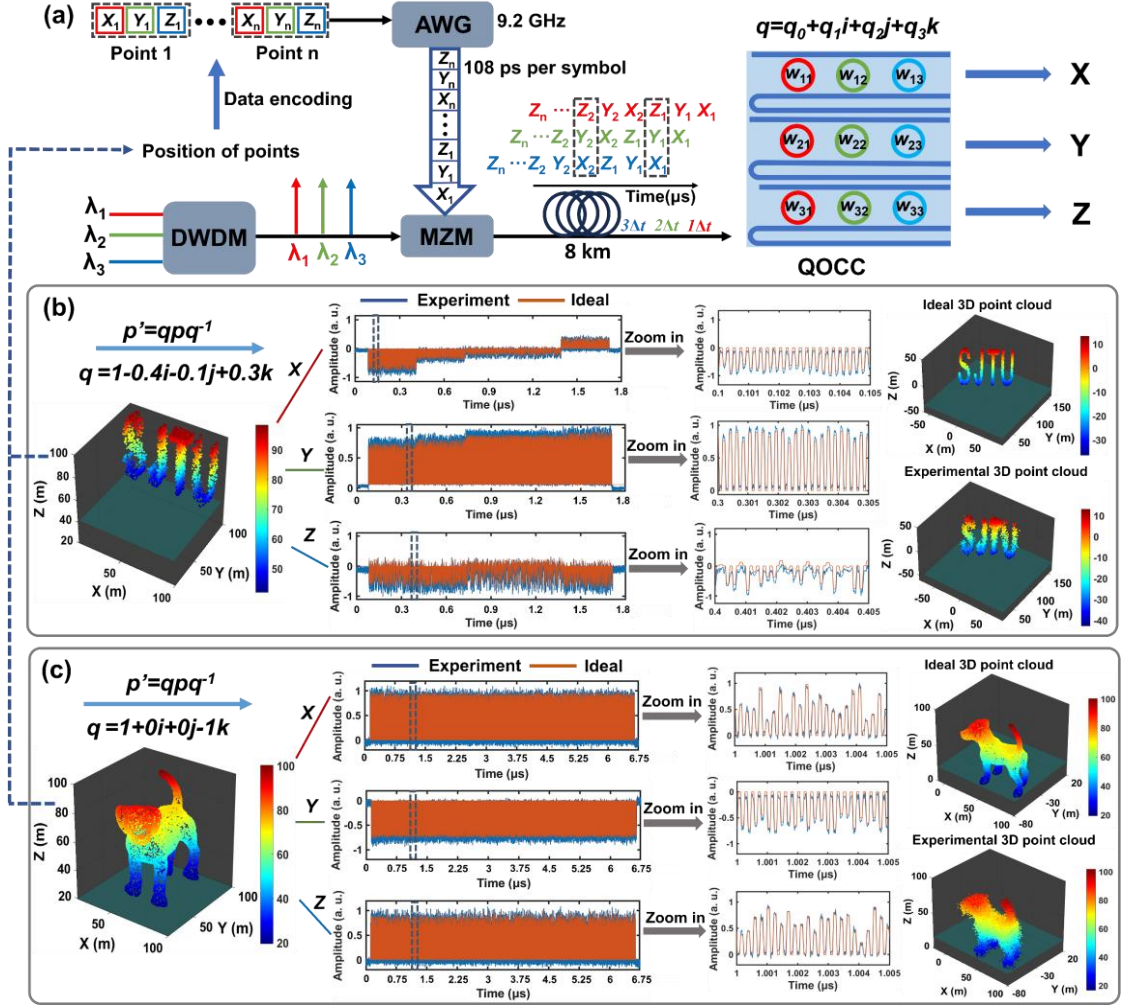

**Fig. 2| 3D point cloud processing by the QOCC. a,** Schematic of the data flow in the quaternion-based 3D point cloud processing experiments. Point cloud coordinate information is encoded onto three optical wavelengths with a modulation speed of 9.2 GHz. Optical signals are multiplexed via a dense DWDM and undergo time-domain wavelength stretching through a 8-km OTDL before being fed into the QOCC for quaternion operations. The computational results are read out via BPDs. **b,c,** Experimental results of the 3D point cloud transformations for the "SJTU" letter model (**b**) and the "Dog" model (**c**). Left panel: the original 3D point cloud image. Middle panel: a comparison between the ideal (red curve) and experimental (blue curve) waveforms of the coordinates of the 3D point cloud after transformations (including an enlarged view). Right panel: a comparative visualization of the output 3D point cloud images after transformations.

Comparing with the 3D point cloud processing schemes employed in conventional electronic chips where the MAC operations are carried out sequentially, the proposed QOCC achieves high parallelism via WDM. Combining with the intrinsic advantages of quaternion computation, it consequently reduces the computational complexity by 2/3 (threefold operations in one propagation). The 3D point cloud processing capability of the QOCC could play a pivotal role in advanced applications such as drone swarm control where real-time and large-scale 3D spatial mapping and drone attitude estimation are mandatory.

**RGB chromatic transformation**

RGB chromatic transformations are a core operation in color image processing, in which the red, green, and blue components are jointly manipulated in 3D color space. By representing RGB pixels as quaternions (i. e., $c = 0 + R\mathbf{i} + G\mathbf{j} + B\mathbf{k}$, where $R$, $G$, and $B$ denote the red, green, and blue channel intensities, respectively), chromatic transformations can be expressed by unit quaternions $q$ ($|q| = 1$) for RGB color space rotations. This method naturally preserves the intrinsic inter-channel correlations and geometric continuity in color space, avoiding the information loss associated with conventional channel-wise processing. Most importantly, it obeys the RGB energy conservation law,

i. e., $R^2 + G^2 + B^2$ is invariant during transformations. Therefore, efficient implementation of RGB chromatic transformation is crucial for high-fidelity color image processing tasks such as color correction, enhancement, and style transformation, particularly in high-dimensional and data-intensive scenarios.

The experimental procedure for the quaternion-based RGB chromatic transformation is illustrated in Fig. 3a. A 512 × 512-pixel saturated color image with RGB components was partitioned into four subsets (512 × 128 × 3 × 4 data blocks) due to the limited oscilloscope memory depth. The input signals $R_i$, $G_i$, $B_i$ were assigned to optical carriers with identical power but distinct wavelengths $\lambda_1$, $\lambda_2$, $\lambda_3$. During signal encoding, the color image pixel values were cyclically formatted as $\mathbf{S} = [R_1, G_1, B_1, R_2... R_n, G_n, B_n]$ according to the pixel order and sequentially fed to the AWG to generate the electrical signals. The signals were then sent into the MZM for electro-optic conversion. The modulated optical signals were launched into the QOCC for quaternion computations. By tuning the resonance of MRRs, the weighting factors for different color channels can be adaptively reconfigured to complete the RGB color space rotation. The signals output from the QOCC were then accumulated and acquired by the BPDs. Similar to the 3D point cloud processing, the redundant data in the output sequence was removed following the criterion described in Supplementary Note 3 before being finally used to generate the transformed color images.

Here we present two examples of RGB chromatic transformations on a color image using the QOCC. The first transformation is a cyclic permutation of RGB channels (i. e., R→G→B→R) represented by a unit quaternion $q = 0.5 + 0.5\mathbf{i} + 0.5\mathbf{j} + 0.5\mathbf{k}$, achieving a 120° rotation in color space. Figure 3b compares the ideal and experimental results of the color image after the 120° rotation. The waveforms of the red, green, and blue channels of the transformed color images are shown in the top panel while a comparative visualization of these two images is displayed in the bottom panel. The good agreement between the ideal and experimental results confirms the high fidelity of the QOCC in performing RGB chromatic transformations. The second transformation is the reverse of the first one, implementing a -120° rotation in color space (i. e., R→B→G→R) represented by the unit quaternion $q = 0.5 - 0.5\mathbf{i} - 0.5\mathbf{j} - 0.5\mathbf{k}$. The ideal and experimental results for this transformation are presented in Fig. 3c, which also demonstrates the high computational fidelity of the QOCC (see details in Supplementary Note 6). For both transformations, the RMSEs between the experimental and ideal waveforms are lower than 0.035, proving that the quaternion-based approach is applicable to any 3D space (no matter whether it is real space or color space). As compared to the conventional electronic processors that process the RGB channels individually, the QOCC leverages the advantages of quaternion computation and the optical parallelism empowered by WDM to accomplish the single-step chromatic transformations, significantly improving the computational efficiency (by 3 times) while successfully preserving the inter-channel correlations.

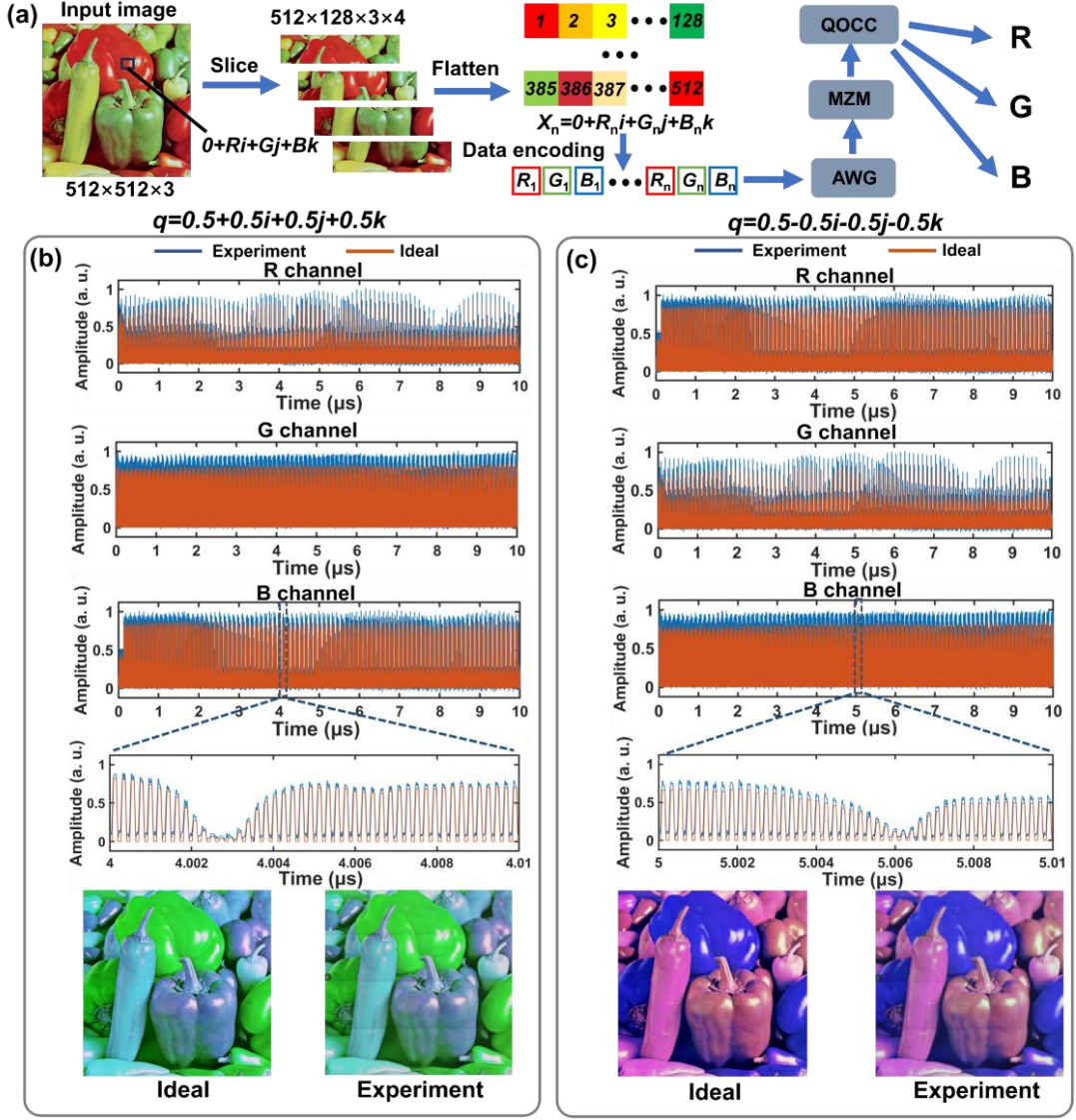

**Fig. 3| RGB chromatic transformations by the QOCC. a,** Schematic of the data flow in the quaternion-based RGB chromatic transformation experiments. **b,c,** Experimental results for the RGB chromatic transformations represented by $q = 0.5 + 0.5\mathbf{i} + 0.5\mathbf{j} + 0.5\mathbf{k}$ (**b**) and $q = 0.5 - 0.5\mathbf{i} - 0.5\mathbf{j} - 0.5\mathbf{k}$ (**c**). Top panel: a comparison between the ideal (red curve) and experimental (blue curve) waveforms of the red, green and blue channels of the transformed color images (including an enlarged view). Bottom panel: a comparative visualization of the ideal and experimental color images after the RGB chromatic transformations.

**Quaternion-valued convolution for color image recognition**

Quaternion convolution (QC) constitutes the core computational operation of QCNNs, which play an essential role in AI tasks such as color image recognition. Traditional real- and complex-valued convolutional neural networks (CNNs) exhibit inherent limitations in color image processing, as convolutional kernels typically aggregate channel-wise outputs independently and thus fail to capture intrinsic inter-channel correlations. In contrast, QCNNs perform convolutions without decomposing the input image into individual RGB channels at the input layer, thereby preserving the interrelated information among channels. More importantly, whereas conventional convolution requires processing three separate color channels, QC treats the color pixel as a unified entity and completes the operation in one step, greatly reducing the computational burden. By fully utilizing the unique properties of quaternion computation, QCNN enhances the ability of neural networks to extract color image features, enabling efficient and high-throughput vision tasks such as color image recognition and segmentation.

To realize the QC operations, we define the pixels of 2D color image as a quaternion matrix $X = [x_{nn'}]_{N \times N}$, where $N$ denotes the image size, and $x_{nn'} = 0 + R_{nn'}\mathbf{i} + G_{nn'}\mathbf{j} + B_{nn'}\mathbf{k}$ is a quaternion.

Similarly, a QC kernel can be written as a quaternion matrix $\boldsymbol{W} = [w_m]_{L \times L}$, where $L$ denotes the kernel size, and $w_m$ is the quaternion element. The QC process can therefore be expressed as

$$\boldsymbol{X} \otimes \boldsymbol{W} = \boldsymbol{Y} = [y_{kk'}]_{(N-L+1) \times (N-L+1)}, \qquad (2)$$

where $\otimes$ represents the QC operation, $y_{kk'}$ is the quaternion-valued pixel in the feature image after convolution. After some cumbersome but straightforward derivations, we find that $y_{kk'}$ is equivalent to (see Supplementary Note 7 for details)

$$y_{kk'} = \sum_{l=1}^{L} \sum_{l'=1}^{L} \begin{bmatrix} w_{m1} & w_{m2} & w_{m3} \\ w_{m3} & w_{m1} & w_{m2} \\ w_{m2} & w_{m3} & w_{m1} \end{bmatrix} x_{(k+l)(k'+l')}, \qquad (3)$$

where $w_{m1}, w_{m2}, w_{m3}$ are the rotation matrix elements derived by substituting $q = w_m$ into Eq. (1).

Next, we show how to use the QOCC to accelerate QCs. The operational principle is illustrated in Fig 4. In our implementation, 2×2 kernels were employed for image convolution, with different kernels realizing diverse convolution functions to improve the effectiveness of feature extraction. The kernels slide on the input image from left to right and then from up to down to implement the computation, following the same procedure as real- and complex-valued convolutions, but now with quaternion arithmetic employed instead. As shown in Fig. 4a, the input 32×32-pixel color image was first sliced into four 31×31-pixel sub-images (Fig. 4b). Then the four sub-images were flattened to generate four quaternion sequences. For instance, the sequences for the first and the fourth sub-images are $\mathbf{S_1} = [x_1, x_2, x_3, ..., x_{31}, x_{33}, x_{34}, ..., x_{989}, x_{990}, x_{991}]$ and $\mathbf{S_4} = [x_{34}, x_{35}, x_{36}, ..., x_{64}, x_{66}, x_{67}, ..., x_{1012}, x_{1023}, x_{1024}]$, respectively. These quaternion sequences were later converted to Data 1 - 4 in RGB format for optical signal encoding (Fig. 4c). Figure 4d illustrates the experimental setup for QC. After multiplexed by a DWDM, the light was equally divided into four paths which were directed to MZM 1 - 4, respectively. The sub-image data was fed to an AWG to drive the MZMs (Data 1 - 4 allocated to MZM 1 - 4, respectively) whose output signals were temporally aligned using OTDLs. The signals were then sent into the QOCC for QC. The QC layer consists of four parallel computing modules (Convolutional module 1 - 4), each independently processing the data stream from the corresponding MZM (MZM 1 - 4). The convolutional kernels were defined by reconfiguring the weighting factors of the MRRs of the QOCC according to Eq. (3). The computational results were detected by BPDs and combined via weighted summation to yield the RGB values of the extracted features *Y* (see Supplementary Note 7 for details).

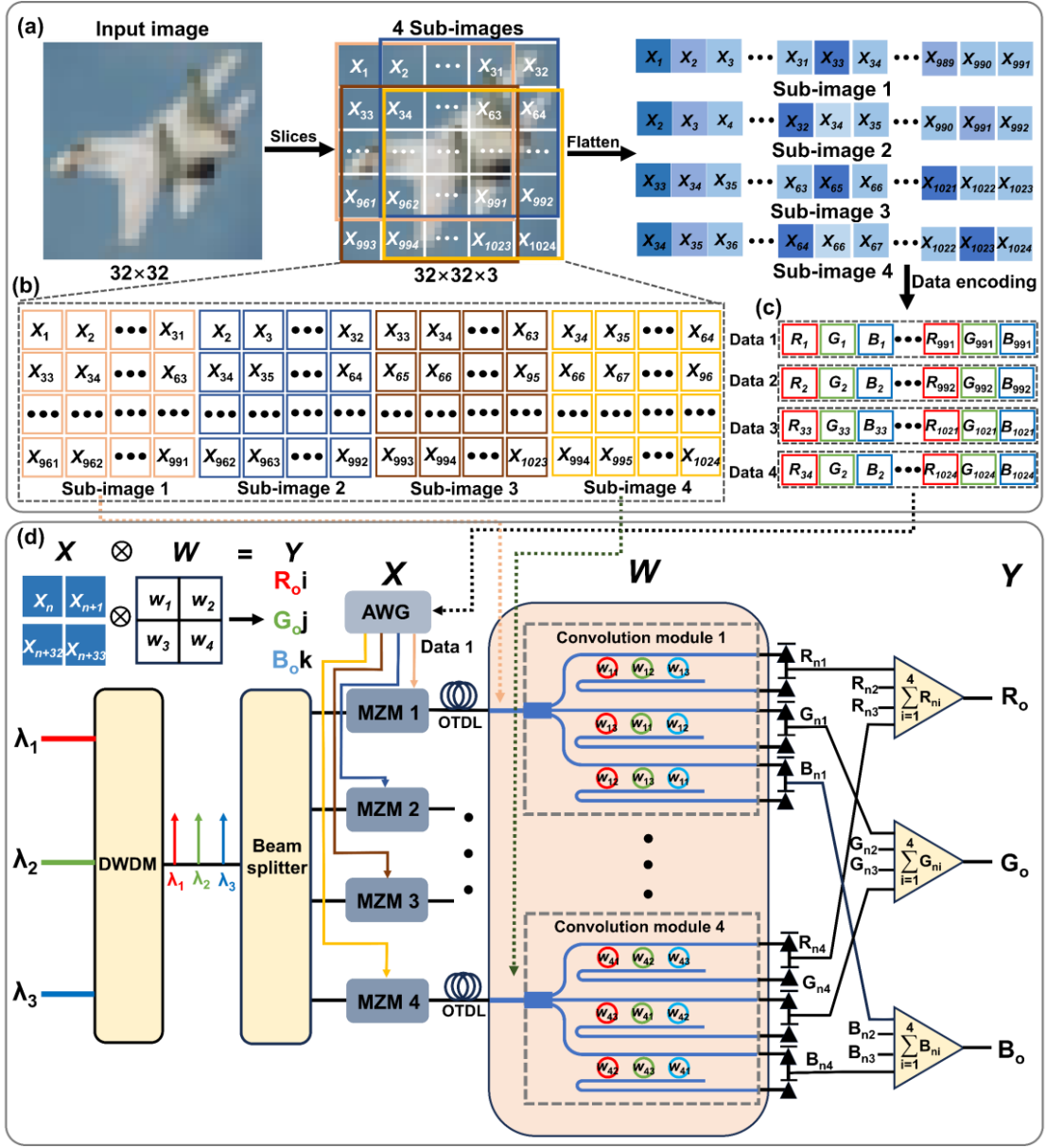

**Fig. 4|** Operational principle of the QC-based feature extraction of the color images. **a,** Pre-processing of the color image. The original image is sliced into four sub-images and then flattened into quaternion sequences for data storage. **b,c,** Data structures of the four sub-images in the quaternion (b) and RGB (c) formats. The flattened pixel sequences serve as the inputs to the AWG. **d,** Schematic of the experimental setup for QC.

To evaluate the performance of the QOCC, we use it in the QCNN to conduct color image recognition experiments on the CIFAR-5 dataset which contains five categories including airplane, automobile, horse, ship, and truck. The employed QCNN consists of an input layer, four convolutional layers and two fully connected layer, as illustrated in Fig. 5a. It was trained on the CIFAR-5 training set comprising 50000 images, with subsequent evaluation performed on a 500-image subset randomly selected from the CIFAR-5 test set. The input layer of the QCNN was 32×32×3 RGB color images, which was then flattened and fed into the first QC layer. The QC layer used a 2×2 convolution kernel to process the 32×32-pixel RGB color images. The kernel W of the first QC layer has the same quaternion elements (i. e, w1 = w2 = w3 = w4 = w) whose transformation matrix R(w) is presented in Fig. 5b. After the initial convolution, 31×31-pixel feature maps of the original images were obtained. These features were then passed through pooling activation before being supplied to subsequent QC layers for further feature extraction[63]. While the theoretical accuracy of QCNN improves with more QC layers, in practice additional layers will contribute to the accumulation of physical errors in photonic processors[64], which include phase errors, amplitude errors, scattering losses and coupling losses. Here, we adopted a four-layer architecture as a balance between the recognition accuracy of QCNN and the experimental errors accumulated across successive layers. The kernels of the other QC layers (QC 2- QC 4) were not fixed and were updated during training. The QCNN employed an end-to-end training framework based on forward-backpropagation[65], with weight optimization implemented with the Adam algorithm (more details can be found in Supplementary Note 8). After all the QC operations, a nonlinear ReLU activation function was applied to transform the 1×9×192 feature maps into 256 feature vectors which were then passed to the fully connected layer. The 1×5 output vector corresponds to the five classes (airplane, automobile, horse, ship, truck), and the predicted class has the maximum value among the five vector elements (see Supplementary Note 9 for more details).

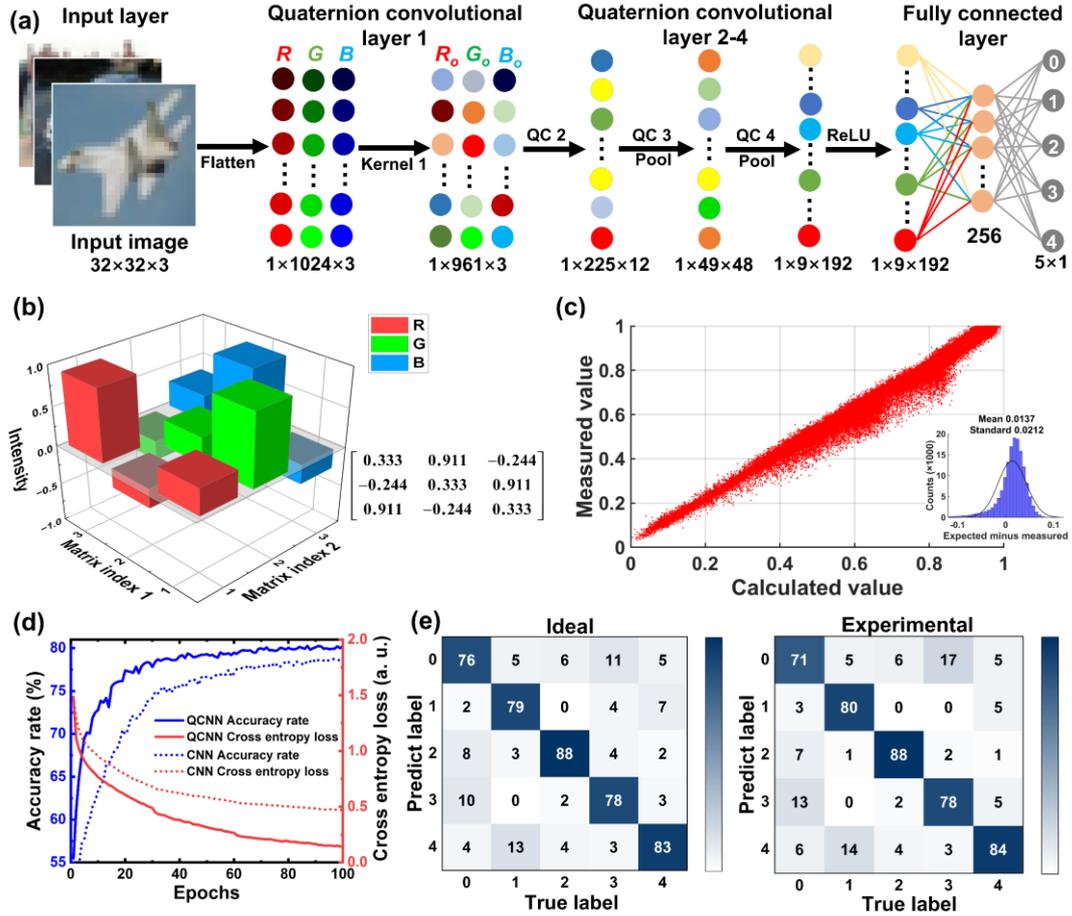

**Fig. 5** | Color image recognition of the CIFAR-5 dataset using the QOCC. **a,** Architecture of the QCNN comprising four activated convolutional layers and two fully-connected layers. QC, Pool and ReLU stand for the quaternion convolutional layer, the pooling layer and the rectified linear unit, respectively. **b,** Weight distribution of the quaternion element within the first QC kernel. **c,** Computational accuracy analysis of the QOCC. The inset histogram shows the computational error distribution with a standard deviation of 0.0212 corresponding to precision of 5 bits. **d,** Recognition accuracies and cross-entropy losses of CNN and QCNN as functions of training epoch. **e,** Comparison between the confusion matrices given by the ideal and experimental QCNNs.

Figure 5c shows the calculation results obtained by the electronic computer (horizontal axis) and the QOCC (vertical axis). To characterize the accuracy of the optical computing method, we calculate the RMSE, mean value, standard deviation and bit precision between the computer- and

QOCC-derived computational results (see Supplementary Note 9 for details). The QOCC exhibits an average RMSE of 0.035, a mean value of ~0.0137, a standard deviation of 0.0212 and 5-bit precision. Figure 5d compares the theoretical recognition accuracies and cross-entropy losses of the QCNN and traditional CNN models as the training epochs progress. The QCNN achieves a saturated accuracy of 82% after 80 training epochs while the CNN reaches 78%. The cross-entropy loss values converge to 0.14 and 0.47 for the QCNN and CNN, respectively. Evidently, the QCNN shows a ~5% improvement in image recognition accuracy over conventional CNNs, which can be attributed to the inter-channel-correlation-preserving property of quaternion computation. Figure 5e presents the ideal confusion matrix derived on computer alongside the experimental matrix obtained via the QOCC for the CIFAR-5 dataset. The optical computing system demonstrates a classification accuracy of ~80.2%, about 2% lower than the ideal prediction. This discrepancy may arise from intrinsic electrical and optical noises, drift of the MRR operating points, amplified spontaneous emission noise from the erbium-doped fiber amplifier, and so on (see more details in Supplementary Note 9). Furthermore, our QOCC implemented on the silicon photonic platform exhibits a superior computing speed compared to conventional electronic chips. Operating at an input data rate of 9.2 Gbaud, the QOCC achieves a computing speed of 9.2 (data rate) × 3 (number of branches per convolutional module) × 4 (number of convolutional modules) × 3 (number of wavelengths) = 331.2 GOPS for QCs. The power consumption for a single QC operation is approximately 108 ps × 3 mW × 9 (number of MRRs per convolutional module) × 4 (number of convolutional modules) = 11.6 pJ. The QOCC enables high-fidelity computation with enhanced processing efficiency, providing a new paradigm for advanced photonic processors that handle computation-intensive tasks such as high-dimensional space transformation and color image recognition.

# Discussion

Quaternion-based computation has manifested itself as a powerful tool in modern AI, showing unique computational advantages in applications such as drone swarm coordination, color image processing and target recognition. However, in traditional electronic chips quaternion computations are generally realized through consecutive MAC operations, leading to substantial computational overhead that hinders large-scale deployments. Here, we propose and experimentally demonstrate a QOCC based on MRR arrays. By leveraging the high parallelism and low latency of optical computing, our design delivers high computational fidelity (RMSE < 0.035) and low computational load (1/3 compared to electronic processors) simultaneously. It also achieves a notably high computing speed of 331.2 GOPS with a record-low energy consumption of 11.6 pJ/operation for QCs, showing superior performance in high-dimensional data processing.

As proof-of-concept demonstrations, we apply the QOCC in three typical quaternion-valued computational tasks including 3D point cloud processing, RGB chromatic transformation and QCNN-based classification of CIFAR-5 dataset. In all these applications, the QOCC produces outcomes with marginal deviations from the ideal electronic-computing benchmarks, but at an accelerated speed and a reduced energy cost. Thanks to the preservation of inter-channel correlations for the RGB color images, the QCNN based on the QOCC successfully achieves a classification accuracy over 80% in color image recognition, outperforming conventional CNNs that process RGB channels separately. Moreover, the photonic processor architecture can be easily scaled up by introducing more MRRs and multiplexing wavelengths. The performance of the computing system is only limited by the experimental setup, e. g., the operational speeds of MZMs and BPDs, the bandwidths of AWG and oscilloscope, and the noise figures of electrical and optical amplifiers. The parallelism of optical computing naturally aligns with the quaternion computation paradigm, enabling complicated quaternion operations such as QCs to be executed in a single light propagation and measurement. With these benefits, our work shows great potential to surpass the computational power of conventional electronic chips in high-dimensional data processing, empowering large-scale computational tasks such as AI training and inference.

Looking ahead, denser integration could be achieved by monolithically incorporating distributed feedback (DFB) lasers, arrayed waveguide gratings, low-loss OTDLs, and high-speed BPDs into a QOCC. Employing phase change materials or magneto-optical materials for non-volatile weight configuration of MRRs would significantly reduce the power consumption of the chip during computation. Overall, the proposed QOCC offers an ideal computational paradigm for next-generation optical computing platforms. With the maturity of integrated photonic technology, the system is expected to feature a smaller footprint, lower power consumption and a higher bandwidth, reaching or exceeding the levels of established electronic computing platforms.

# Methods

**Experimental setup**

The QOCC was fabricated by the MPW foundry (CUMEC, China) using the standard CSiP180Al line PDK. Titanium nitride (TiN) was deposited on top of the waveguides to serve as the microheaters. Isolation trenches were created for suppressing the thermal crosstalk between adjacent devices. The details of the QOCC are provided in Supplementary Note 2. To characterize the fabricated devices, a tunable continuous wave (CW) laser (Santec TSL-770) and a photodetector (PD, Santec MPM-210) were employed. The polarization of light from the tunable laser was first adjusted by a fiber polarization controller (PC, Thorlabs FPC032). Then the TE-polarized light was coupled into and out of the chip through grating couplers. An optical power meter was used for optical calibration. The wavelength of the laser was swept from 1500 nm to 1600 nm to measure the transmission spectra of the devices.

The experimental setup for quaternion computations was built with commercial optoelectronic devices and the QOCC. The DFB (IDPHOTONICS) laser array generated the light at three wavelengths of 1550.8 nm, 1552.4 nm and 1554 nm, with adjustable output powers ranging from 10 to 20 dBm. The three wavelength channels were modulated by a 40-GHz MZM (iXblue) which converts the 65-GSa/s electrical waveforms from the AWG (KEYSIGHT M9502A) to the optical signals. The multi-channel programmable power supply (TIME-TRANSFER) provides 64 tunable voltage channels (in the range of 0 - 10 V) for the weight configuration of the MRR arrays. After passing through the QOCC, the optical signals were converted to temporal electrical signals through BPDs (FINISAR BPDV2150RQ) with an operational bandwidth exceeding 50 GHz. The time-domain waveforms of the signals were then sampled using a high-speed oscilloscope (TELEDYNE LECROY) with a sampling rate of 80 Ga/s.

**Weight configuration of the QOCC**

The weight configuration of the QOCC was implemented by an automated system. In the automated scanning algorithm, the control program was implemented using Python 3.12, where the socket and PyVISA libraries were employed to establish serial communications with the multi-channel voltage source and the oscilloscope, respectively. The system interfaced with the multi-channel voltage source through the UDP protocol and concurrently acquired measurement signals from the oscilloscope via the SCPI protocol. During operation, each MRR was subjected to a linear voltage sweep, with the oscilloscope concurrently recording the corresponding weight at every voltage point. The algorithm then determined the optimal weights for the MRRs (see Supplementary Note 10 for more details).

**Acknowledgements**

The work was supported in part by the National Key Research and Development Program of China (2023YFB2905503(C. L.)), the National Natural Science Foundation of China (62475146(L. S.), and 62341508(Y. S.)), and the Open Fund of Key Laboratory of Photonic Technology for Integrated Sensing and Communication, Ministry of Education (L. S.).

**Competing interests**

The authors declare no competing interests.

**Data availability**

The data that support the findings of this study are provided in the Supplementary Information/Source Data file. Source data are provided with this paper.